%% file: main.tex
\documentclass[a4paper]{jpconf}
\usepackage{graphicx}
\usepackage[labelfont=bf]{caption}
\usepackage{subcaption}
\usepackage{amsmath}
\usepackage{amssymb}

\usepackage{xurl}
\usepackage{tikz}
\usetikzlibrary{quantikz}
\usepackage{bm} %Bold math  symbol
\usetikzlibrary{shapes,arrows}
\usetikzlibrary{positioning}
\usetikzlibrary{shadows}

\begin{document}
\title{Running the Dual-PQC GAN on noisy simulators and real quantum hardware}

\author{Su Yeon Chang$^{1,2}$, Edwin Agnew$^3$, El\'ias F. Combarro$^4$, Michele Grossi$^1$, Steven Herbert$^{3,5}$ and Sofia Vallecorsa$^1$}
\address{$^1$ CERN Openlab, 1211 Geneva 23, Switzerland}
\address{$^2$ Institute of Physics, Ecole Polytechnique F\'ed\'erale de Lausanne (EPFL), 1015 Lausanne, Switzerland}
\address{$^3$ Cambridge Quantum, Cambridge, Terrington House, 13–15 Hills Road, CB2 1NL, United Kingdom}
\address{$^4$ Department of Computer Science, University of Oviedo, 33003 Oviedo, Asturias, Spain}
\address{$^5$ Department of Computer Science and Technology, University of Cambridge, Cambridge CB2 1TN, United Kingdom}
\ead{steven.herbert@cambridgequantum.com, su.yeon.chang@cern.ch}
\begin{abstract}
In an earlier work \cite{DualPQC}, we introduced dual-Parameterized Quantum Circuit (PQC) Generative Adversarial Networks (GAN), an advanced prototype of a quantum  GAN. We applied the model on a realistic High-Energy Physics (HEP) use case: the exact theoretical simulation of a calorimeter response with a reduced problem size.
This paper explores the dual-PQC GAN for a more practical usage by testing its performance in the presence of different types of quantum noise, which are the major obstacles to overcome for successful deployment using near-term quantum devices. The results propose the possibility of running the model on current real hardware, but improvements are still required in some areas.

\end{abstract}

\section{Introduction}
\input{1Intro}

\section{Dual-PQC GAN}
\input{2DualPQC}

\section{Hyperparameter scan on noise model}
\input{3NoiseModel}

\section{Results on real hardware}
\input{4RealHardware}

\section{Conclusions}
\input{5Conclusion}
\ack{This project is supported by CERN’s Quantum Technology Initiative. We acknowledge the use of IBM Quantum, and Amazon Web Services' quantum computing service Amazon Braket for implementing work described in this article. The views expressed are only those of the authors, and do not reflect the official policy or position of IBM, the IBM Quantum team, Amazon team or Amazon Web Services. One of the authors was supported in part under grant PID2020-119082RB-C22 funded by MCIN/AEI/10.13039/501100011033 and grant AYUD/2021/50994 funded by Gobierno del Principado de Asturias (Spain).}

\section*{References}
\bibliographystyle{iopart-num}
\bibliography{biblio.bib}

\end{document}

%% file: 1Intro.tex
\label{sec:Introduction}

Quantum computing emerges as a promising technique to complement the traditional `classical' computing, thanks to its potential to speed up computations or solve problems that classical algorithms cannot address \cite{QuantumSupremacy}. However, current Noisy Intermediate-Scale Quantum (NISQ) devices suffer from intrinsic noise from different sources which yield biased results. Quantum Machine Learning (QML) via Variation Quantum Algorithms (VQA) is one of the algorithms which can be successfully simulated with a reasonable circuit depth in the presence of noise \cite{VQA}.  Still, the influence of noise is not negligible, which is compounded by the  vanishing gradient problem \cite{NIBP}, therefore further research is required to improve the performance of algorithms.  

In order to understand the impact of quantum noise in QML, our work specifically focuses on Dual-Parameterized Quantum Circuit (PQC) Generative Adversarial Networks (GAN), which are a new prototype of quantum GANs characterized by a classical discriminator and two quantum generators that take the form of PQCs. In our earlier work \cite{DualPQC}, we have demonstrated that the dual-PQC GAN is able to imitate reduced size pixelated images of calorimeter outputs in High-Energy Physics (HEP), but only in the absence of noise.

This paper investigates the training of dual-PQC GAN in the presence of quantum noise. We start by briefly summarizing the architecture of dual-PQC GAN in Section \ref{sec:dualPQC}. In Section \ref{sec:NoiseModel}, we test the dual-PQC GAN on noise models with  two-qubit gate errors only, which are the dominating factor in current architectures. In particular, the impact of different training hyperparameters is investigated within a wider range of errors with respect to that of current real hardware. Finally, Section \ref{sec:RealHardware} presents the results obtained both on superconducting and trapped-ion quantum hardware for the optimal hyperparameters found in noisy simulations. Ultimately, our work  aims to provide a global overview of the effect of different types of noise in the training of dual-PQC GAN and suggests realistic solutions to provide model convergence.

%% file: 2DualPQC.tex
\label{sec:dualPQC}
This section summarizes the architecture of dual-PQC GAN model, which is a new type of quantum GAN that we introduced in a previous study \cite{DualPQC}. It is a hybrid qGAN architecture which has one classical discriminator and two parameterized quantum circuits, PQC1 and PQC2, sharing the role of the generator. 

Consider a training set $X \subset \mathbb{R}^{2^n}$ of $N = 2^n$ pixel images following a certain distribution.  PQC1, with $n_1$ qubits, learns the probability distribution over image samples by measuring its output state to produce $n_1$ bits.  This bit string is then used to initialise PQC2 (of $n_2$ qubits) with the corresponding computational basis, $\ket{i} \in \mathbb{R}^{2^{n_1}}$. By repeatedly measuring $n$ output qubits of PQC2, a probability distribution is constructed over $2^{n}$ computational basis states and \textit{interpreted} as an image, $\mathcal{I}_i$, of size $2^n$. 
The classical discriminator takes the training set and the
images generated by PQC2, and it classifies them into real and
fake.  Ultimately, the dual-PQC GAN model can generate $2^{n_1}$ images of
size $2^n$. 

 \tikzstyle{Generator2} = [rectangle, draw, fill=green!20, 
text width=4.6em, text centered, rounded corners, minimum height=1.3cm]
\tikzstyle{Discriminator2} = [rectangle, draw, fill=blue!20, 
text width=6.8em, text centered, rounded corners, minimum height=1cm]
\tikzstyle{input2} = [rectangle, draw, fill= white, 
text width=0.3em, minimum height = 2.3em, align= center,
copy shadow={draw, shadow xshift=1mm, shadow yshift=-1mm} ]
\tikzset{meter/.append style={draw, inner sep=4, rectangle, font=\vphantom{A}, minimum width=16, line width=.36,
    path picture={\draw[black] ([shift={(.04,.12)}]path picture bounding box.south west) to[bend left=50] ([shift={(-.04,.12)}]path picture bounding box.south east);\draw[black,-latex] ([shift={(0,.04)}]path picture bounding box.south) -- ([shift={(.12,-.04)}]path picture bounding box.north);}}}
    
\tikzstyle{line} = [draw, -latex']
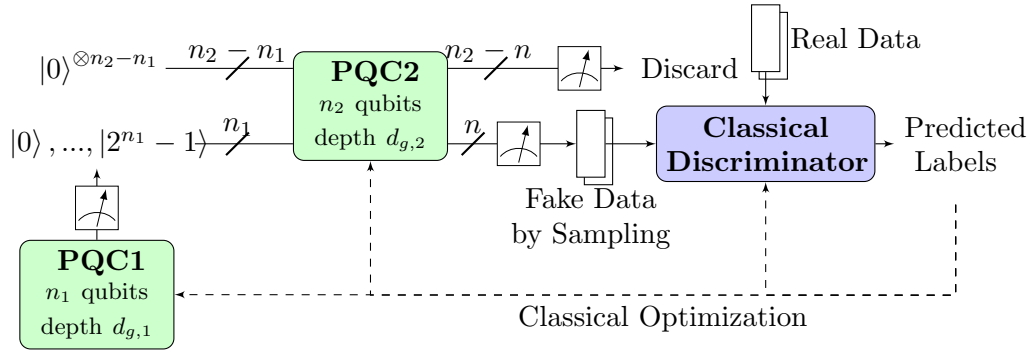
\begin{figure}[h]
  \begin{center}
    \begin{tikzpicture}[align=center,node distance = 2cm, auto]
    % Place nodes

    \draw (6.5,-0.5) node[input2, label = below:{
      Fake Data \\ by Sampling}] (FakeData) {\phantom{input}};
    \node [Discriminator2, right of = FakeData, node distance=2.3cm] (Disc) {\bf   Classical \\ Discriminator}; 
    \node [input2, above of = Disc, node distance=1.4cm, label = right:{Real Data}] (RealData) {\phantom{input}};
    \node [text width = 3.5em, right of = Disc, node distance = 2.5cm, minimum height = 1.6cm](PredictedLabel) {Predicted Labels};
    \node [text width = 3.8em](Discard) at (7.8,0.5) {Discard};
    \node [rectangle, text width=6em, minimum height = 1em, align= center] (Basis) at (0,-0.5) {${\ket{0},...,\ket{2^{n_1}-1}}~~~~$};
    \node [rectangle, text width=4em, minimum height = 1em, align= center] (Zeros) at (0,0.5) {$\ket{0}^{\otimes n_2 - n_1}~~~~$};
    \node [Generator2, below of = Basis, node distance = 2cm] (PQC1) { { \textbf{PQC1}} \\  \footnotesize $n_1$ qubits \\ depth $d_{g,1}$};
    \node [Generator2] (PQC2) at (3.6,0) {{ \textbf{PQC2}} \\ \footnotesize $n_2$ qubits \\ depth $d_{g,2}$};
    
    \draw[thick] (5.1, 0.35) -- (5.4, 0.65);
    \draw[thick] (4.8, -0.65) -- (5.1, -0.35);
    
    \draw[thick] (1.7, 0.35) -- (2, 0.65);
    \draw[thick] (1.7, -0.65) -- (2, -0.35);
    
    \node[text width = 5em] at (5.15,0.72) {$n_2 - n$};
    \node[text width = 1em] at (4.95,-0.28) {$n$};
    
    \node[text width = 5em] at (1.85,0.72) {$n_2 - n_1$};
    \node[text width = 1em] at (1.85,-0.28) {$n_1$};
    % Draw edges
    \path [line] (PQC1) -- (Basis);
        \draw (Basis.east)  |- (PQC2.west |-Basis.east); 
        \draw (Zeros.east)  |- (PQC2.west |-Zeros.east); 
    \path [line] (PQC2.east|-FakeData.west) |- (FakeData);
    \path [line] (PQC2.east|-Discard.west) |- (Discard);
    \path [line] (FakeData.east) -- (Disc);
    \path [line] (Disc) -- (PredictedLabel);
    \path [line] (RealData) -- (Disc);
    \path [line, style = dashed] (PredictedLabel.south) -- ++(0,-1.2cm)  -| (Disc.south);
    \path [line, style = dashed] (PredictedLabel.south) -- ++(0,-1.2cm)  -|node [pos = 0.25, below] (TextNode) {Classical Optimization}  (PQC2.south);
    \path [line, style = dashed] (PredictedLabel.south) -- ++(0,-1.2cm)  -- (PQC1.east);
    
    \node[meter, above of = PQC1, node distance = 1.15cm](meter){};
    \node[meter, left of = Discard, node distance = 1.45cm](meter){}; 
    \node[meter, left of = FakeData, node distance = 0.95cm](meter){};      
    \end{tikzpicture}
  \end{center}
  \caption{\label{fig2}Schematic diagram of dual-PQC GAN to reproduce images of $2^n$ pixels.}
\end{figure}

In Ref \cite{DualPQC}, we apply Dual-PQC GAN on a realistic HEP use-case to reproduce calorimeter outputs of a reduced size, which are interpreted as pixelated images. In order to facilitate the comparison between real and fake images, we classify the training set into 4 classes via K-means clustering and average over each class. Note that this classification is simply for the comparison, and the whole set of images is used for the training.  

In the simulations, the performance of the model is quantified with two metrics: 
\begin{enumerate}
    \item \textbf{Relative entropy}, $D_{KL}(\mathcal{I}_{mean} \| \tilde{\mathcal{I}}_{mean})$, between the average of the real images, $\tilde{\mathcal{I}}_{mean}$, and the generated images, $\mathcal{I}_{mean}$, with  $
        D_{KL}(p\| q) = \sum_{j} p(j)\log\frac{p(j)}{q(j)}    
   $
    \item \textbf{\textit{Individual} relative entropy}, the mean of the minimum relative entropy for each of the generated images with respect to the real images: $
        D_{KL, ind} = \frac{1}{2^{n_1}}\sum_{i = 0}^{2^{n_1} -1}\min_{j} D_{KL}(\tilde{\mathcal{I}}_{j}\| \mathcal{I}_{i})     
    $.
\end{enumerate}

%% file: 3NoiseModel.tex
\label{sec:NoiseModel}
Before running the model on real hardware, we search for the \textit{optimal} hyperparameters using grid search over different factors: PQC1 and PQC2 learning rate, discriminator learning rate, and decay rate. The hyperparameter scan was executed with the \textit{Qiskit} noise model which mimics the noise of a real quantum device, using the dual-PQC model with $n_1 = 2$, $n_2 = 4$, $d_1 = 2$ and $d_2 = 5$. As the readout error is negligible compared to the two-qubit gate error in this configuration, only the latter is included in this noise model. 

\begin{figure}[h]
\centering
\begin{subfigure}{0.31\textwidth}
    \includegraphics[width = \textwidth]{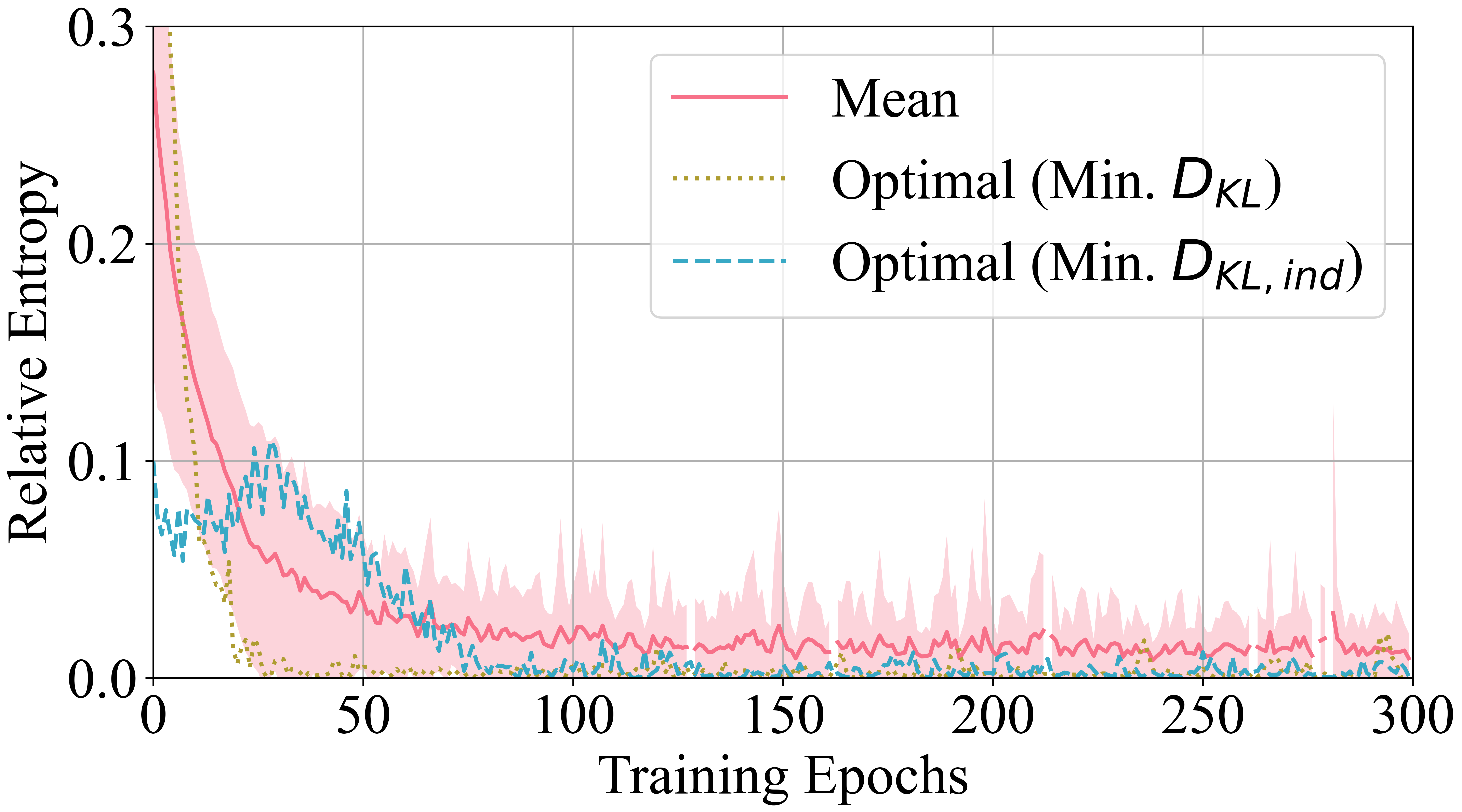}
    \caption{$p=0$}
    \label{fig:scan_no_noise}
\end{subfigure}
\begin{subfigure}{0.31\textwidth}
    \includegraphics[width = \textwidth]{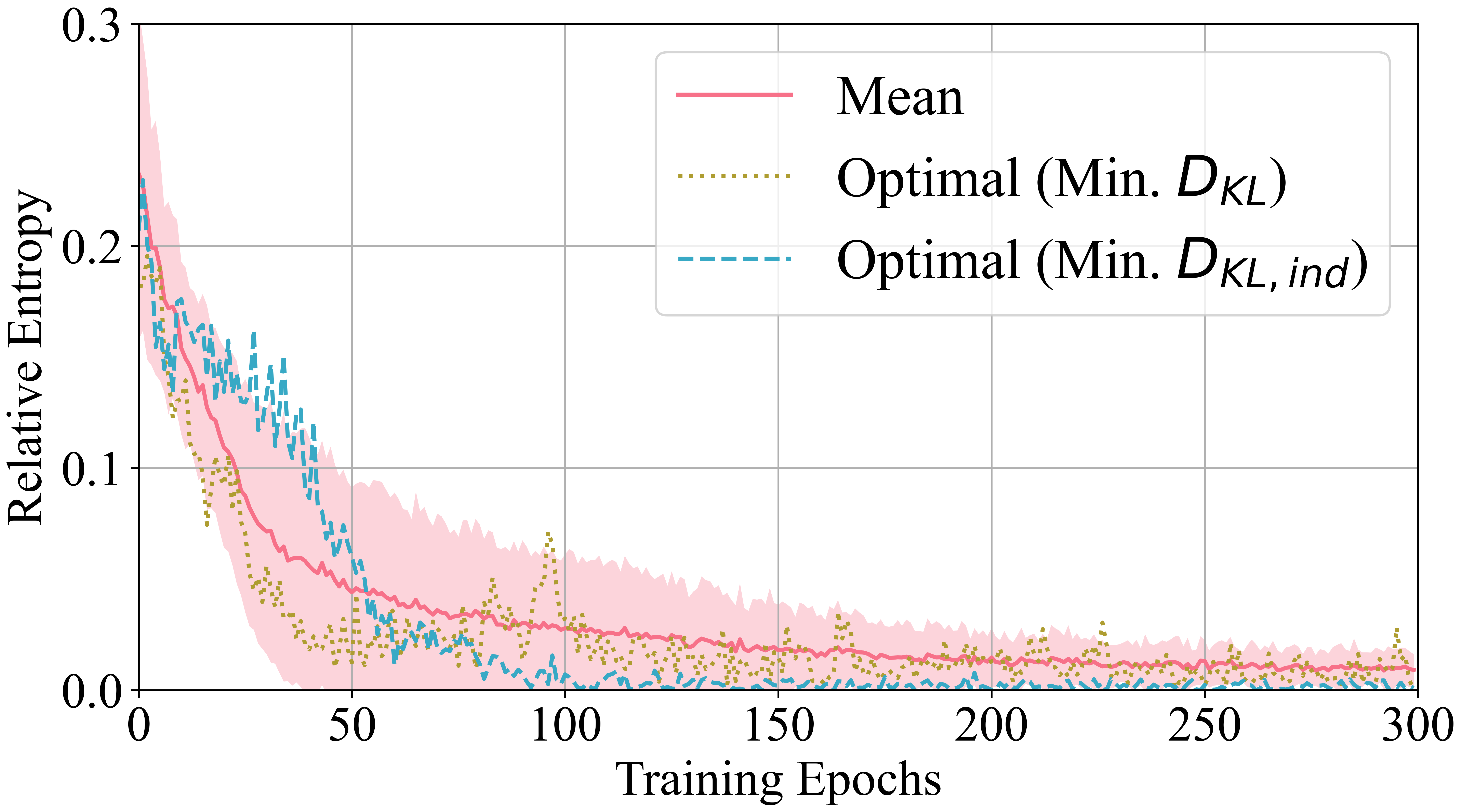}
    \caption{$p=0.02$}
    \label{fig:scan_p002}
\end{subfigure}
\begin{subfigure}{0.31\textwidth}
    \includegraphics[width = \textwidth]{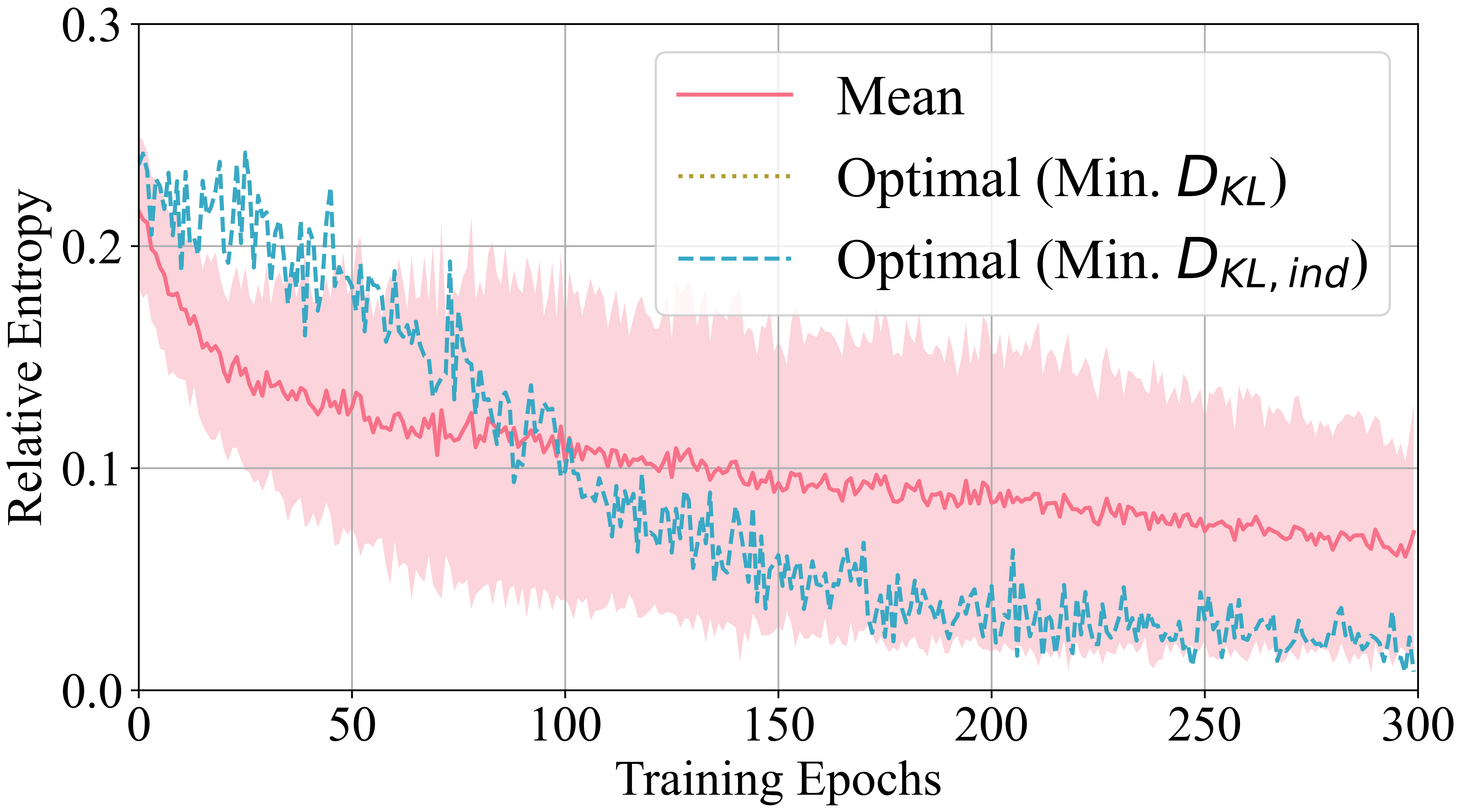}
    \caption{$p=0.04$}
    \label{fig:scan_p004}
\end{subfigure}
\caption{The relative entropy obtained from the hyperparameter scan with different two-qubit gate error, $p$. The results with the lowest $D_{KL}$ and $D_{KL, ind}$ at the end of the training are also displayed with two plots overlapping on (c). Note that all the tests where the losses diverge are excluded from the plot.}
\label{fig:scan}
\end{figure} 

\begin{figure}[h]
\centering
\begin{subfigure}{0.3\textwidth}
    \includegraphics[width = \textwidth]{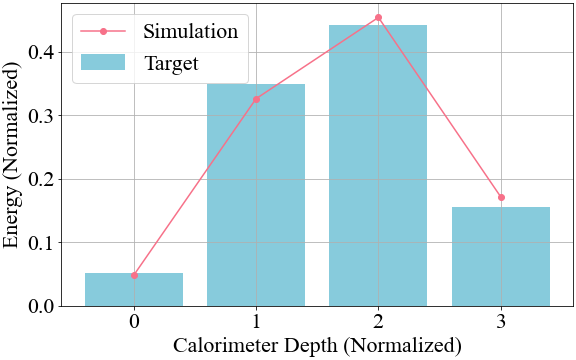}
    \caption{$p=0$}
    \label{fig:mean_image_best1_no_noise}
\end{subfigure}
\begin{subfigure}{0.3\textwidth}
    \includegraphics[width = \textwidth]{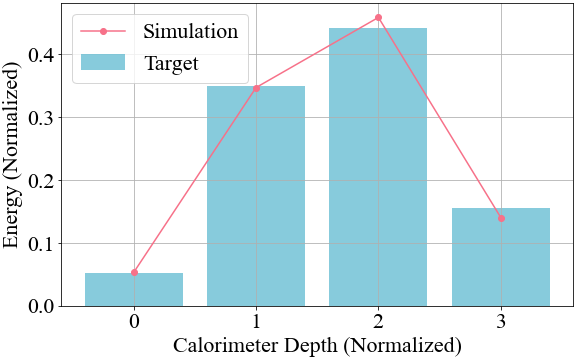}
    \caption{$p=0.02$}
    \label{fig:mean_image_best1_p002}
\end{subfigure}
\begin{subfigure}{0.3\textwidth}
    \includegraphics[width = \textwidth]{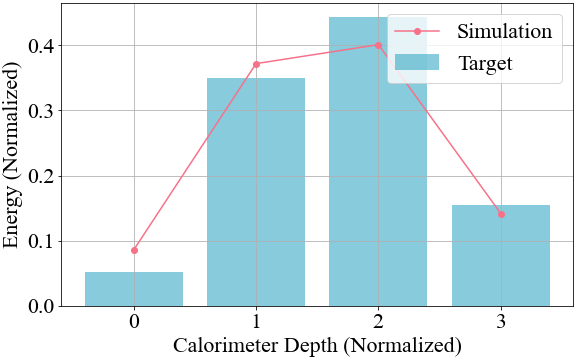}
    \caption{$p=0.04$}
    \label{fig:mean_image_best1_p004}
\end{subfigure}
\begin{subfigure}{0.3\textwidth}
    \includegraphics[width = \textwidth]{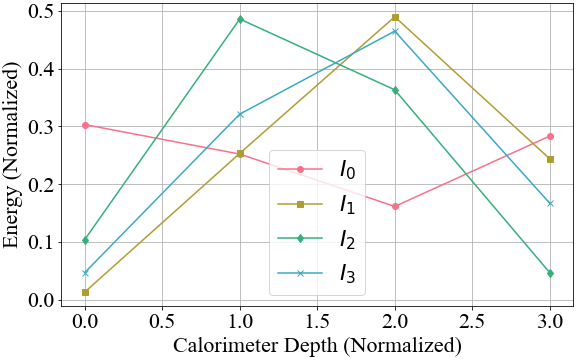}
    \caption{$p=0$}
    \label{fig:ind_image_best1_no_noise}
\end{subfigure}
\begin{subfigure}{0.3\textwidth}
    \includegraphics[width = \textwidth]{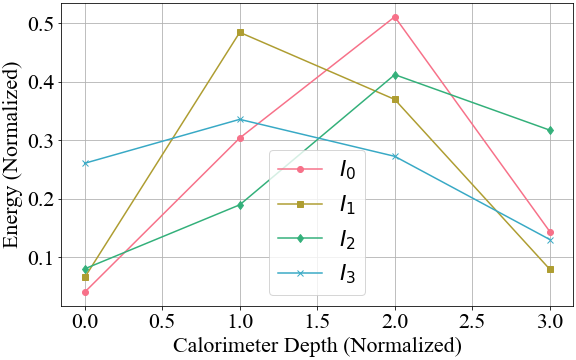}
    \caption{$p=0.02$}
    \label{fig:ind_image_best1_p002}
\end{subfigure}
\begin{subfigure}{0.3\textwidth}
    \includegraphics[width = \textwidth]{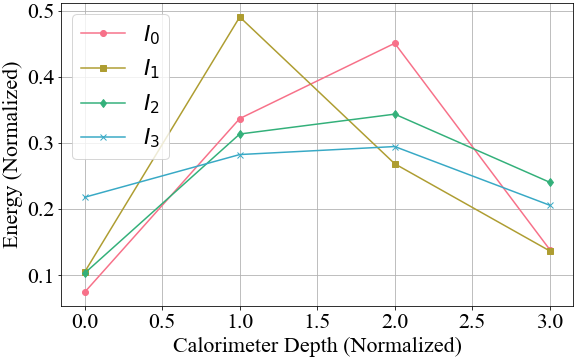}
    \caption{$p=0.04$}
    \label{fig:ind_image_best1_p004}
\end{subfigure}
\caption{The mean image (a,b,c) and individual images (d,e,f) obtained at the end of the Dual-PQC GAN training with the optimal hyperparameters, which give the lowest $D_{KL,ind}$ for different two-qubit gate errors, $p$. }
\label{fig:scan_images}
\end{figure} 
As shown in Figure \ref{fig:scan}, the standard deviation of the relative entropy increases as the two-qubit gate error $p$ increases. But at the same time, we consistently get hyperparmeters which make the training converge in all cases, proving that the model can be trained even in the presence of noise under the optimal choice of hyperparameters. The convergence in mean and individual images obtained in the presence of different two-qubit gate errors shown in Figure \ref{fig:scan_images}  also emphasizes this fact.

%% file: 4RealHardware.tex
\label{sec:RealHardware}
\subsection{Inference on real hardware}
Before training the dual-PQC GAN on real quantum hardware, we start by testing the inference of the pre-trained model in order to check the impact of real quantum noise, including two-qubit gate errors and readout errors, on the generation of images. This also allows us to compare the performance of currently accessible quantum device technologies: superconducting chips from \textit{IBM} \cite{IBMQ_calibration}, and ion-trap machines from \textit{IONQ} \cite{IONQ}.  Regarding IBM chips, qubits with the lowest error rates are explicitly chosen from the IBM Quantum Lab while the IONQ machines were accessed via Amazon Web Services (AWS) Braket.

\begin{figure}[h]
\centering
\begin{subfigure}{0.23\textwidth}
    \centering
    \includegraphics[width = \textwidth]{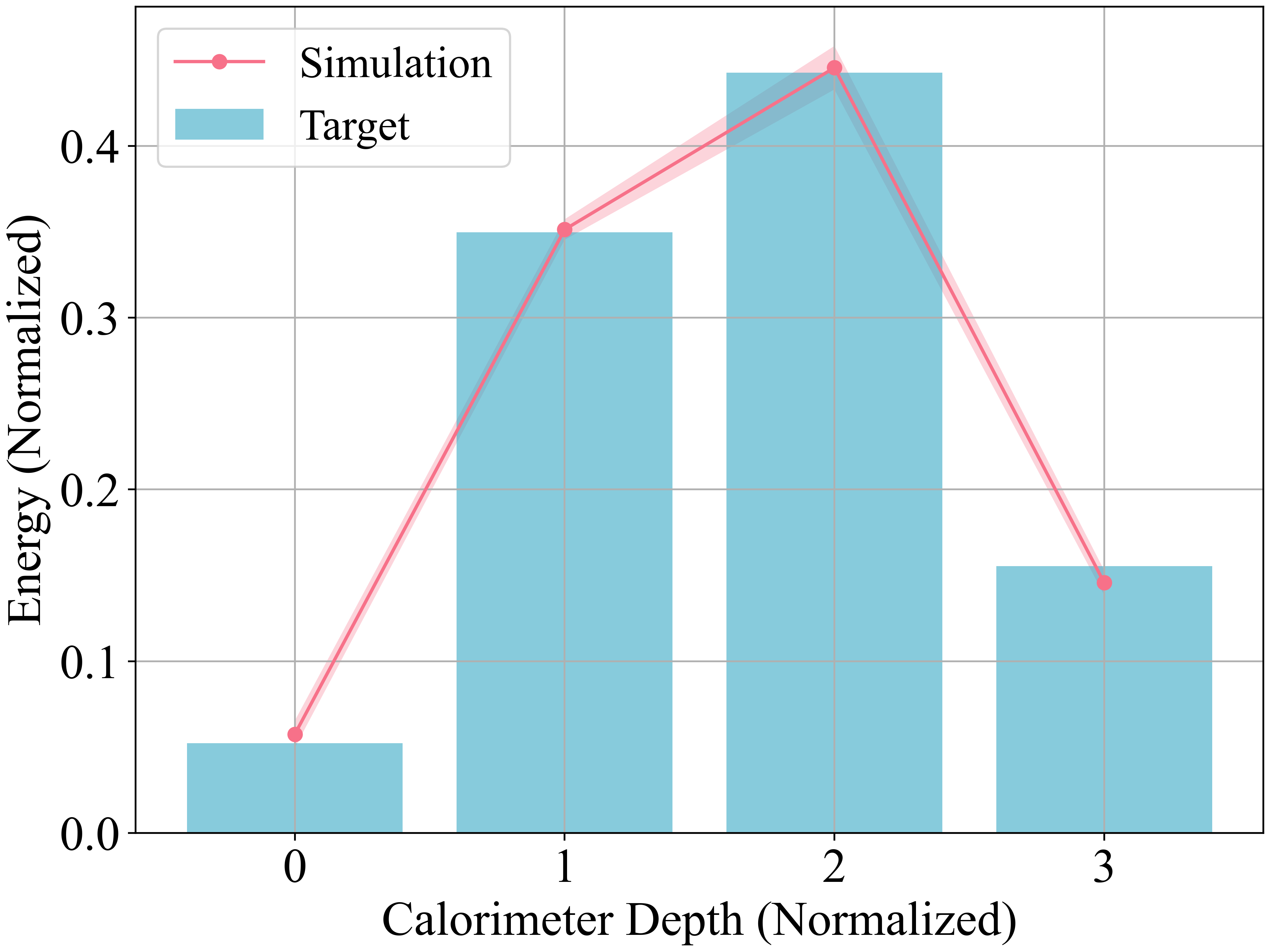}
    \caption{}
    \label{fig:mean_image_ibmq_jakarta}
\end{subfigure}
\begin{subfigure}{0.23\textwidth}
    \centering
    \includegraphics[width = \textwidth]{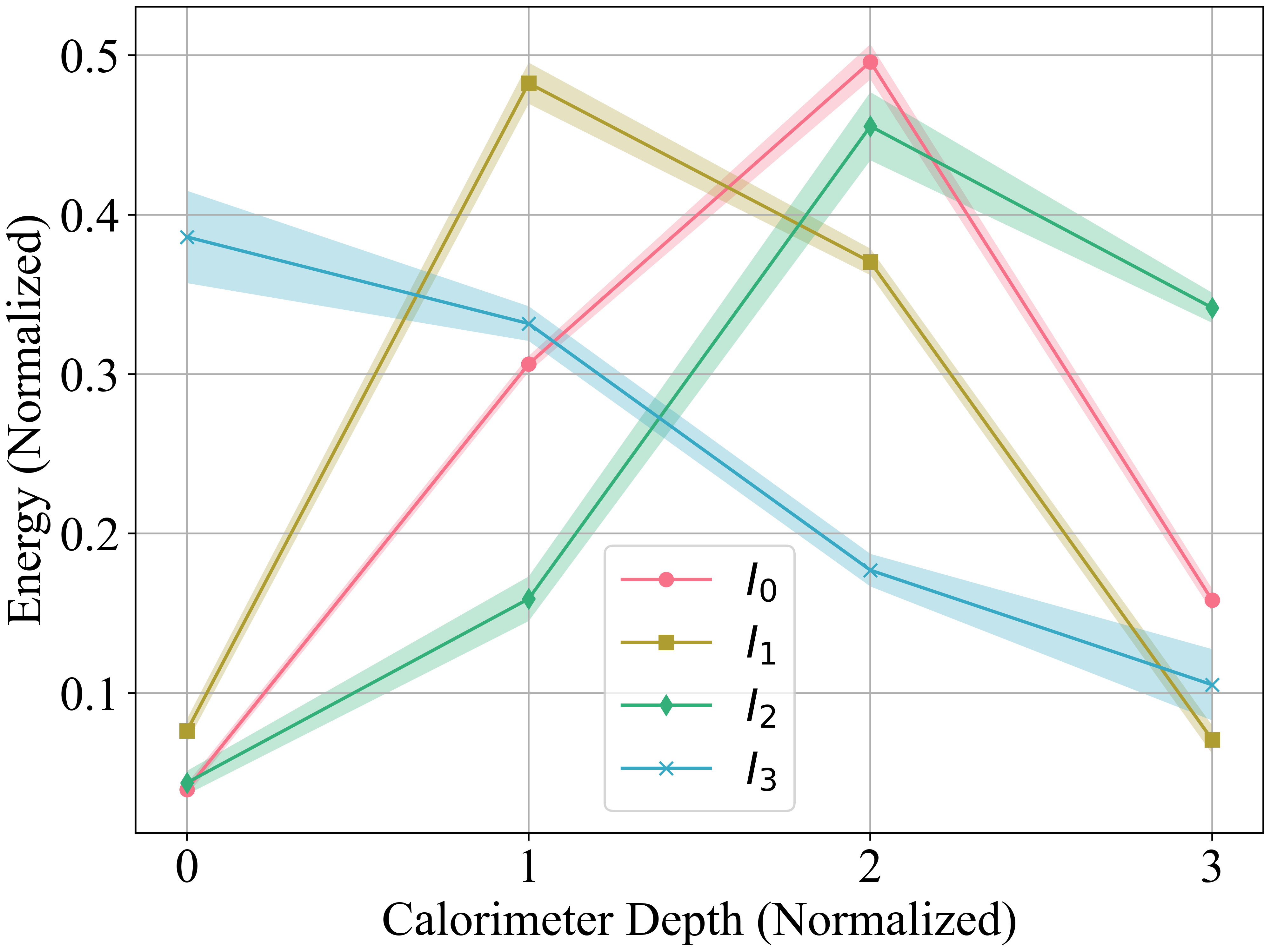}
    \caption{}
    \label{fig:images_ibmq_jakarta}
\end{subfigure}
\begin{subfigure}{0.23\textwidth}
    \centering
    \includegraphics[width = \textwidth]{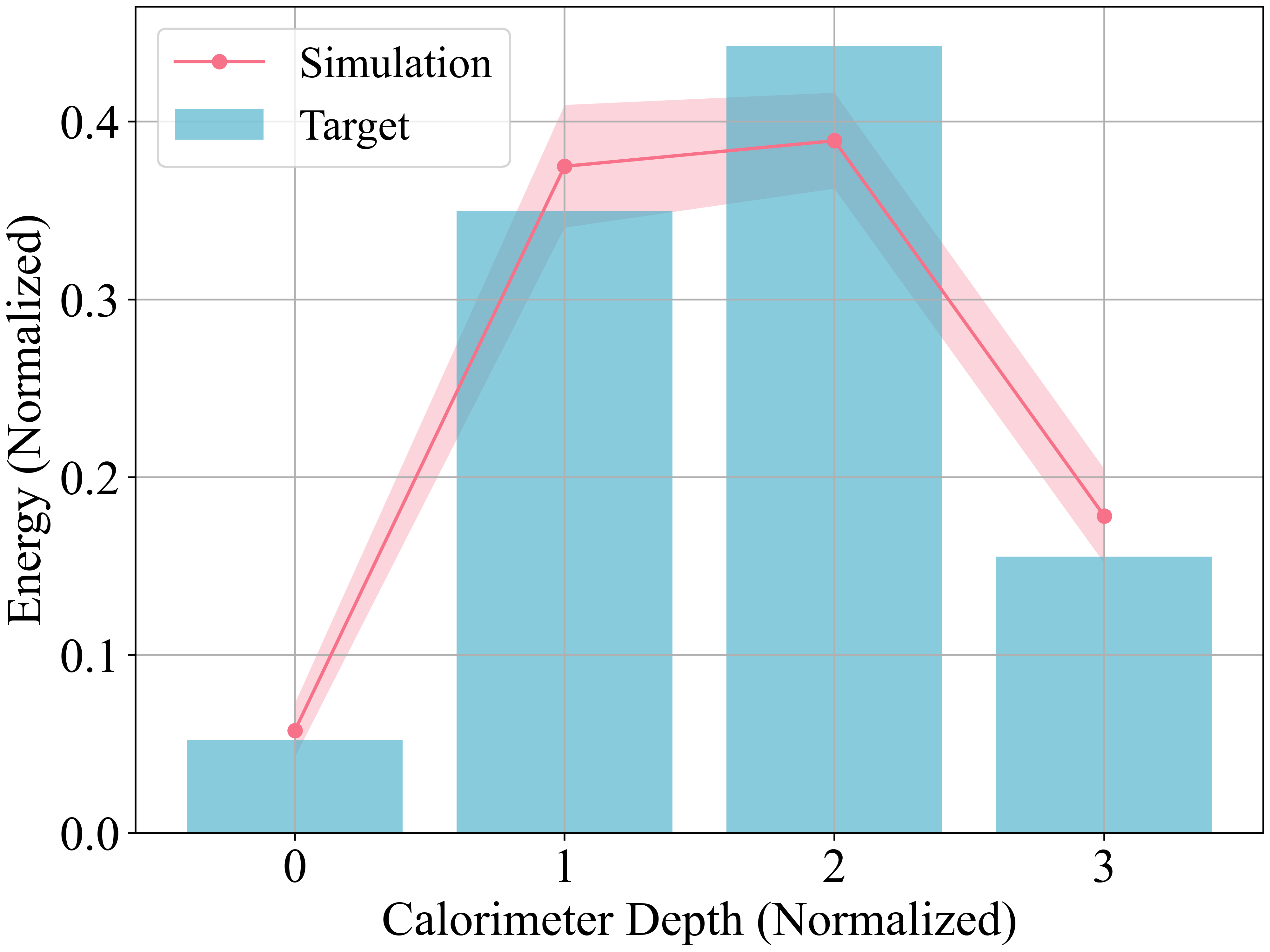}
    \caption{}
    \label{fig:mean_image_ionq}
\end{subfigure}
\begin{subfigure}{0.23\textwidth}
    \centering
    \includegraphics[width = \textwidth]{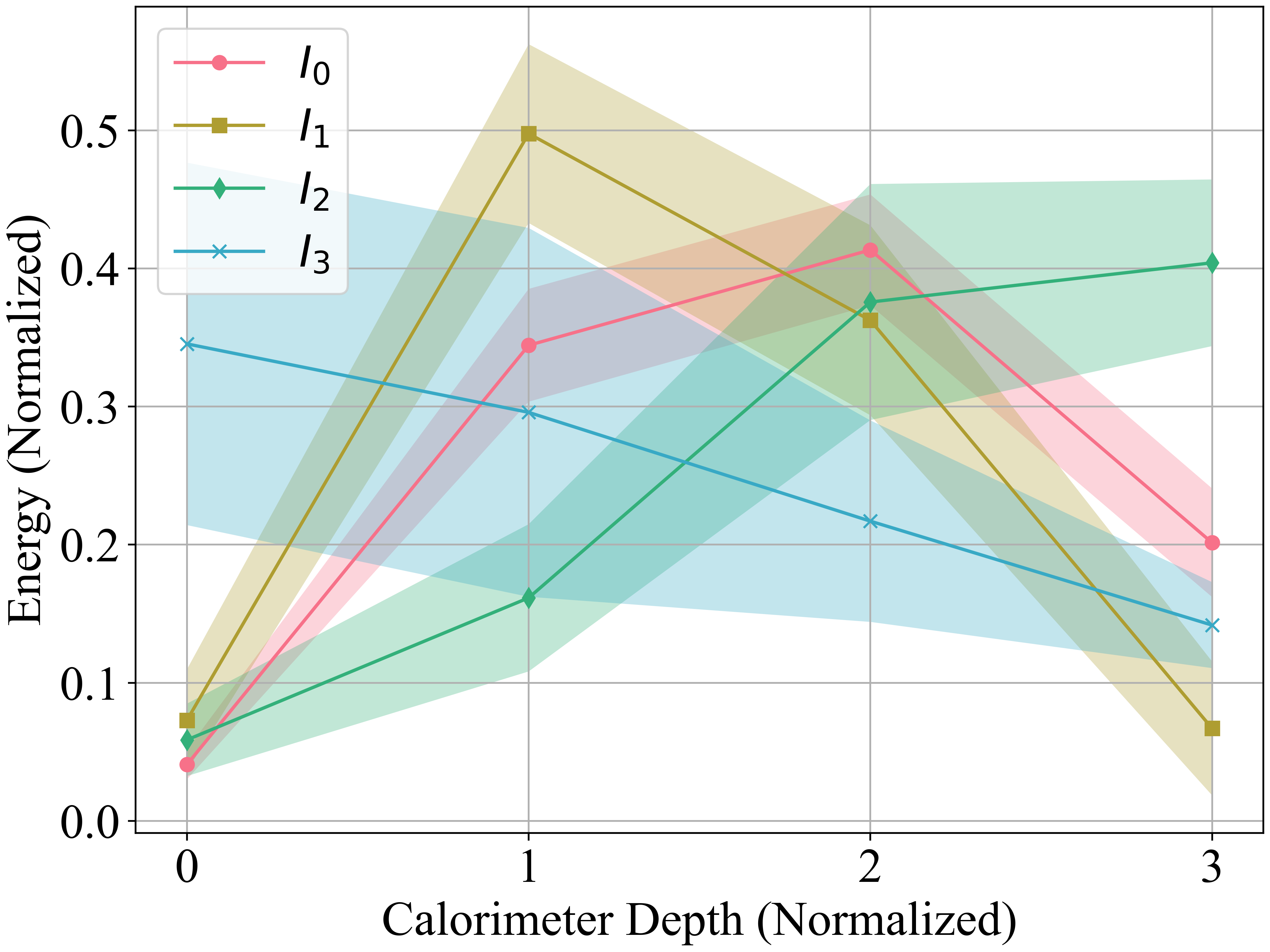}
    \caption{}
    \label{fig:images_ionq}
\end{subfigure}
\caption{Mean (a,c) and individual images (b,d) obtained by inference test on ibmq\_jakarta (a,b) and IONQ (c,d). }
\label{fig:inference}
\end{figure}

Figure \ref{fig:inference} presents the mean and individual images obtained from inference tests on real quantum devices. The pre-trained weights are taken from the test which reproduced Figure \ref{fig:mean_image_best1_p002} and \ref{fig:ind_image_best1_p002}. Looking at the low standard deviation suggest the feasibility of dual-PQC training for superconducting chips, while more studies are required with currently accessible trapped-ion devices. This prediction is also confirmed  in Table \ref{tab:inference} which summarizes the error rates of each devices and the result of inference tests. 

\begin{table}[h]
\caption{$D_{KL}$ and $D_{KL, ind}$ averaged over 20 inference tests on different quantum hardware and their error rates \cite{IBMQ_calibration}. The results obtained with the two-qubit gate error only noise model is also displayed for comparison.}
\label{tab:inference}
\begin{center}
    \lineup
    \begin{tabular}{ccccc}
    \br
      Device & Readout error & CX error & $D_{KL}$ ($\times 10^{-2}$) & $D_{KL, ind}$ ($\times 10^{-2}$) \\
      \mr
      Noise Model & NULL & $2.00\cdot10^{-2}$ &  $0.07 \pm 0.04$ & $5.54 \pm 0.04$ \\
      ibmq\_jakarta & $2.80\cdot10^{-2}$ & $1.37\cdot10^{-2}$ &  $0.14 \pm 0.14$ & $6.49 \pm 0.54$ \\
      ibm\_lagos & $1.15\cdot10^{-2}$ & $5.58\cdot10^{-3}$ &   $0.26 \pm 0.11$ & $6.92 \pm 0.71$ \\
      ibmq\_casablanca & $2.61 \cdot10^{-2}$ & $4.58\cdot10^{-2}$  & $4.03 \pm 1.08$ & $6.58 \pm 0.81$ \\
      ibm\_perth & $2.34 \cdot10^{-2}$& $1.68\cdot10^{-2}$  & $0.71 \pm 0.19$ & $10.03 \pm 0.65$ \\
      IONQ & NULL & $1.59\cdot10^{-2}$  & $1.24 \pm 0.74$ & $10.10 \pm 5.62$ \\
      \br
    \end{tabular} 
\end{center}
\end{table}

\subsection{Training on real hardware}

\begin{figure}[h]
\centering
\begin{subfigure}{0.3\textwidth}
    \centering
    \includegraphics[width = \textwidth]{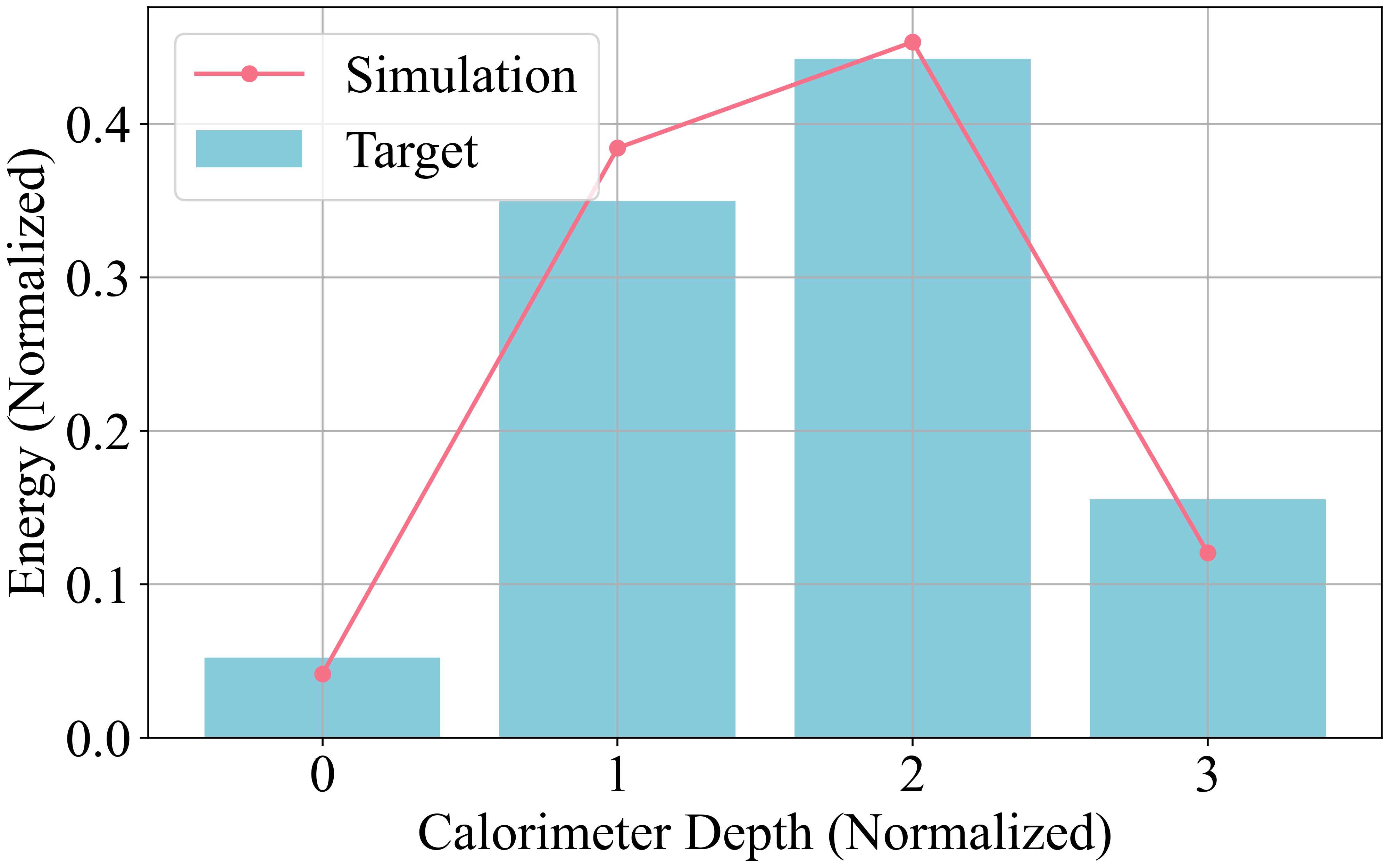}
    \caption{}
    \label{fig:mean_image_lagos}
\end{subfigure}
\begin{subfigure}{0.3\textwidth}
    \centering
    \includegraphics[width = \textwidth]{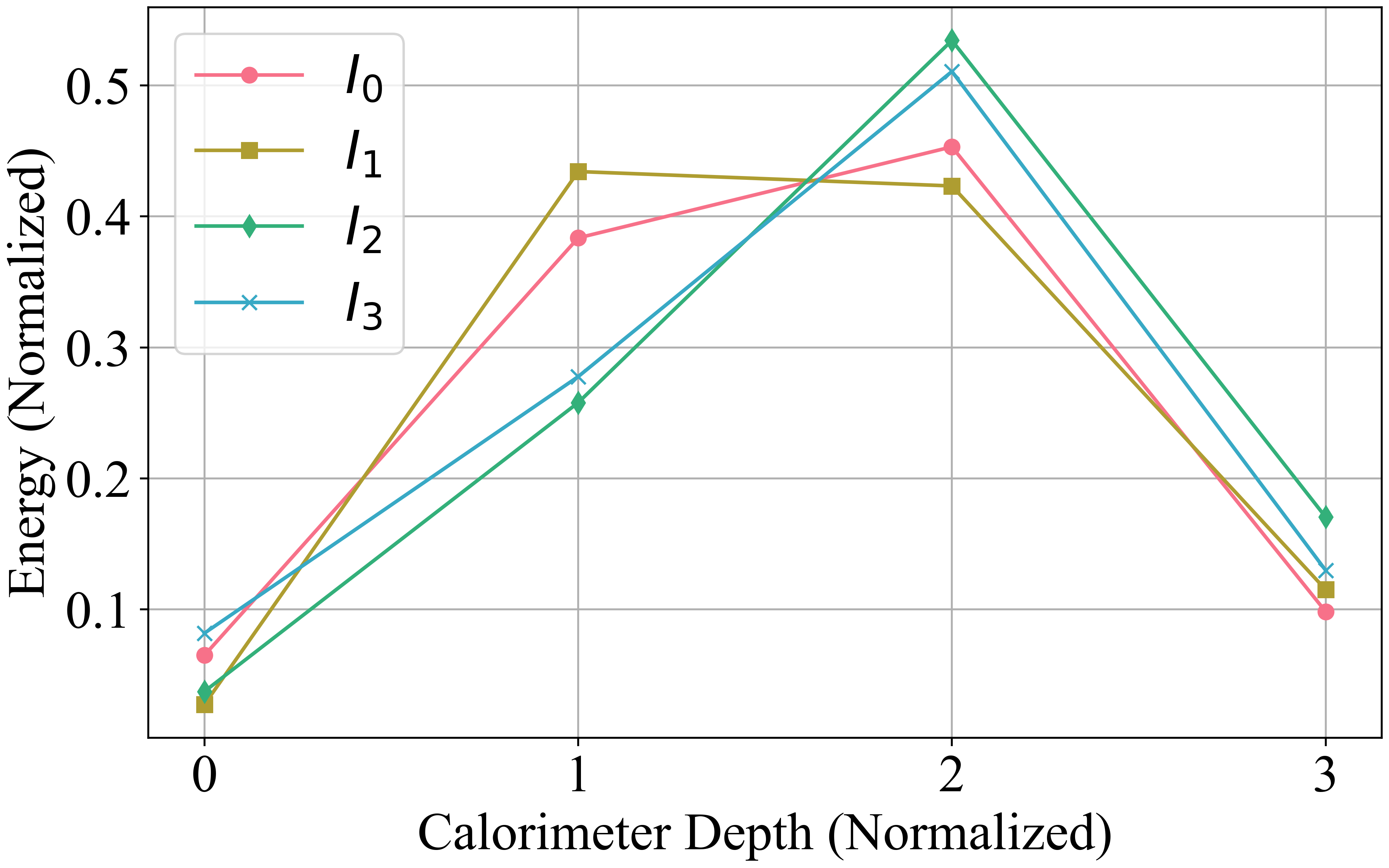}
    \caption{}
    \label{fig:images_lagos}
\end{subfigure}
\begin{subfigure}{0.3\textwidth}
    \centering
    \includegraphics[width = \textwidth]{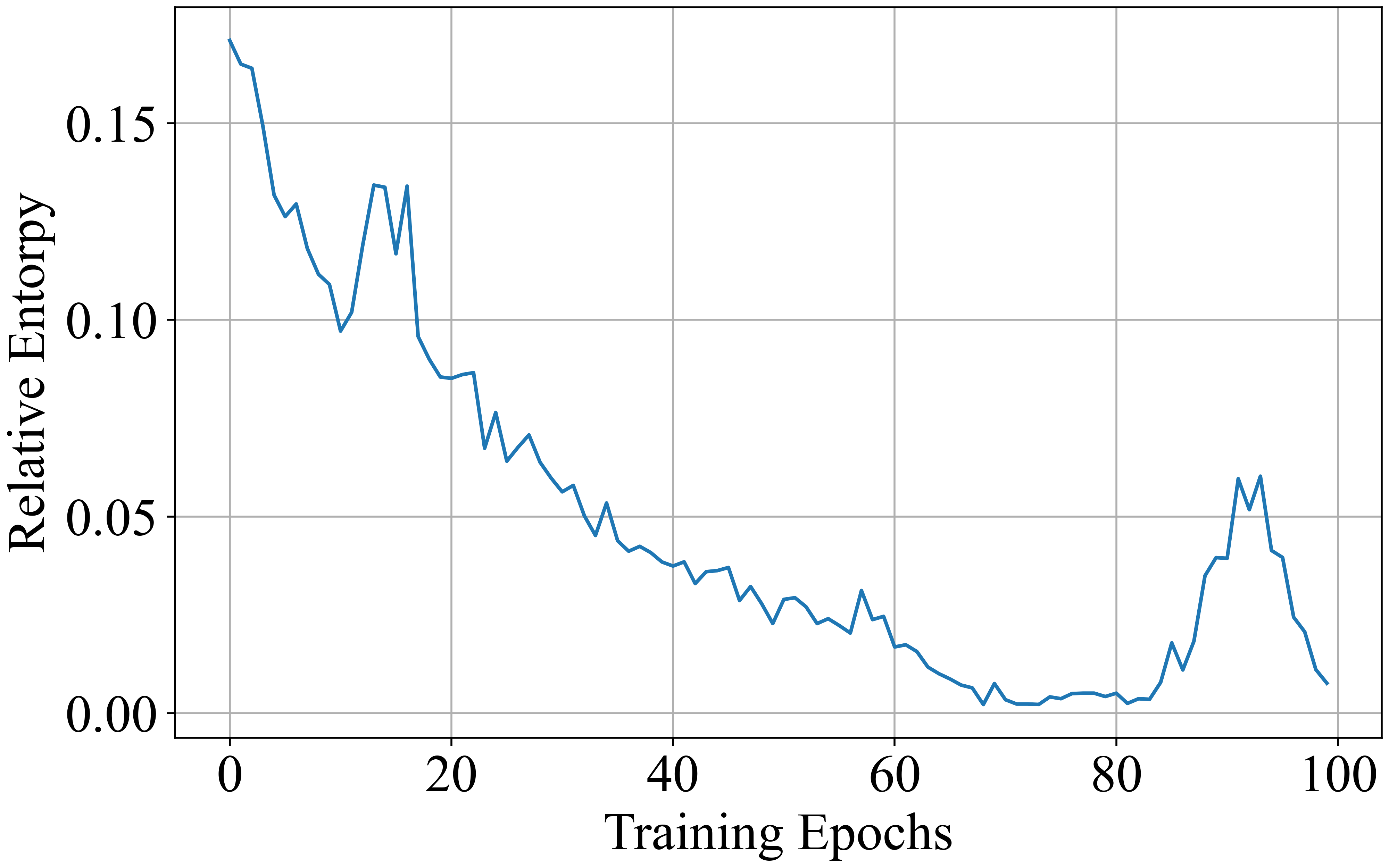}
    \caption{}
    \label{fig:rel_entr_lagos}
\end{subfigure}
\begin{subfigure}{0.3\textwidth}
    \centering
    \includegraphics[width = \textwidth]{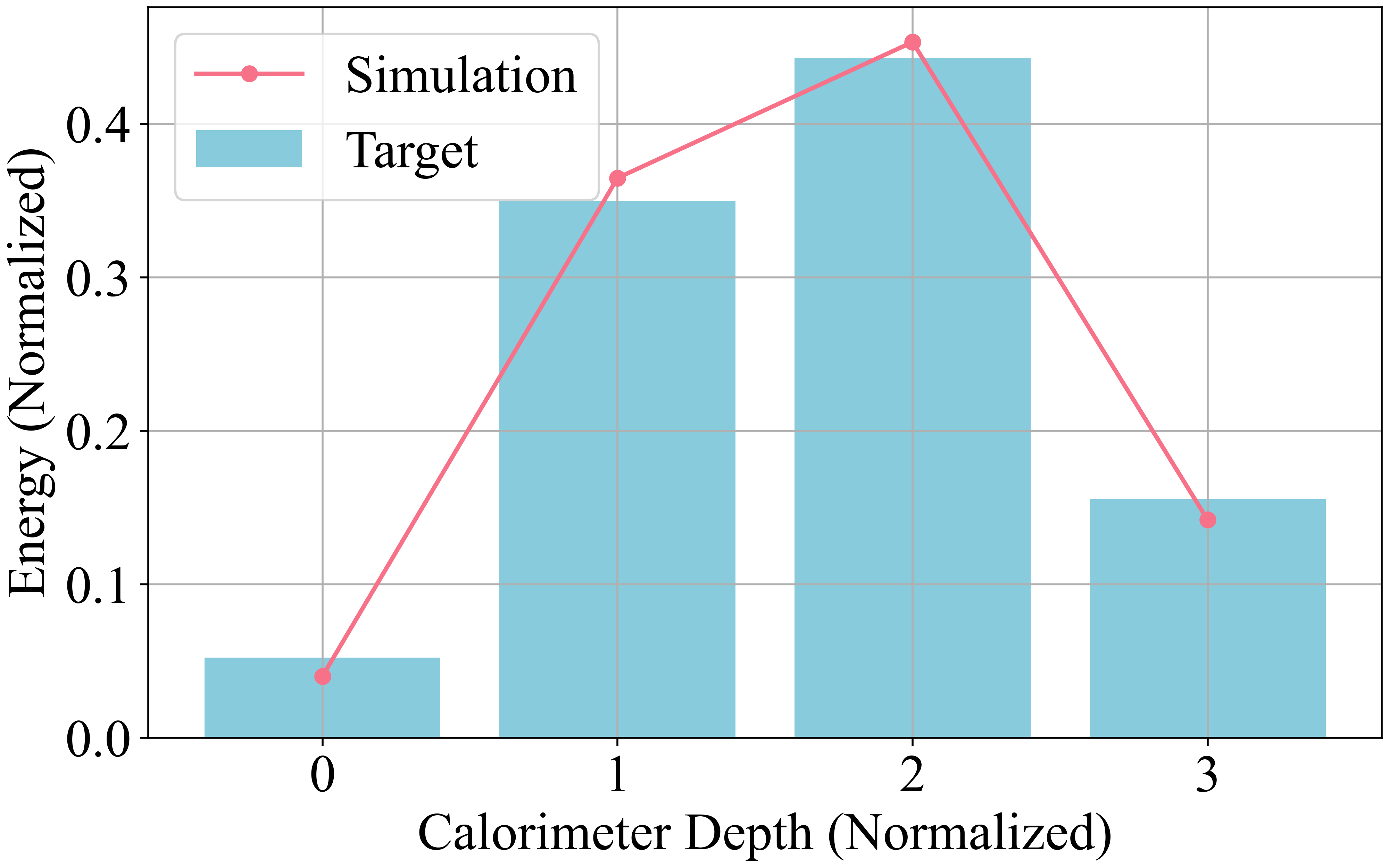}
    \caption{}
    \label{fig:mean_image_perth}
\end{subfigure}
\begin{subfigure}{0.3\textwidth}
    \centering
    \includegraphics[width = \textwidth]{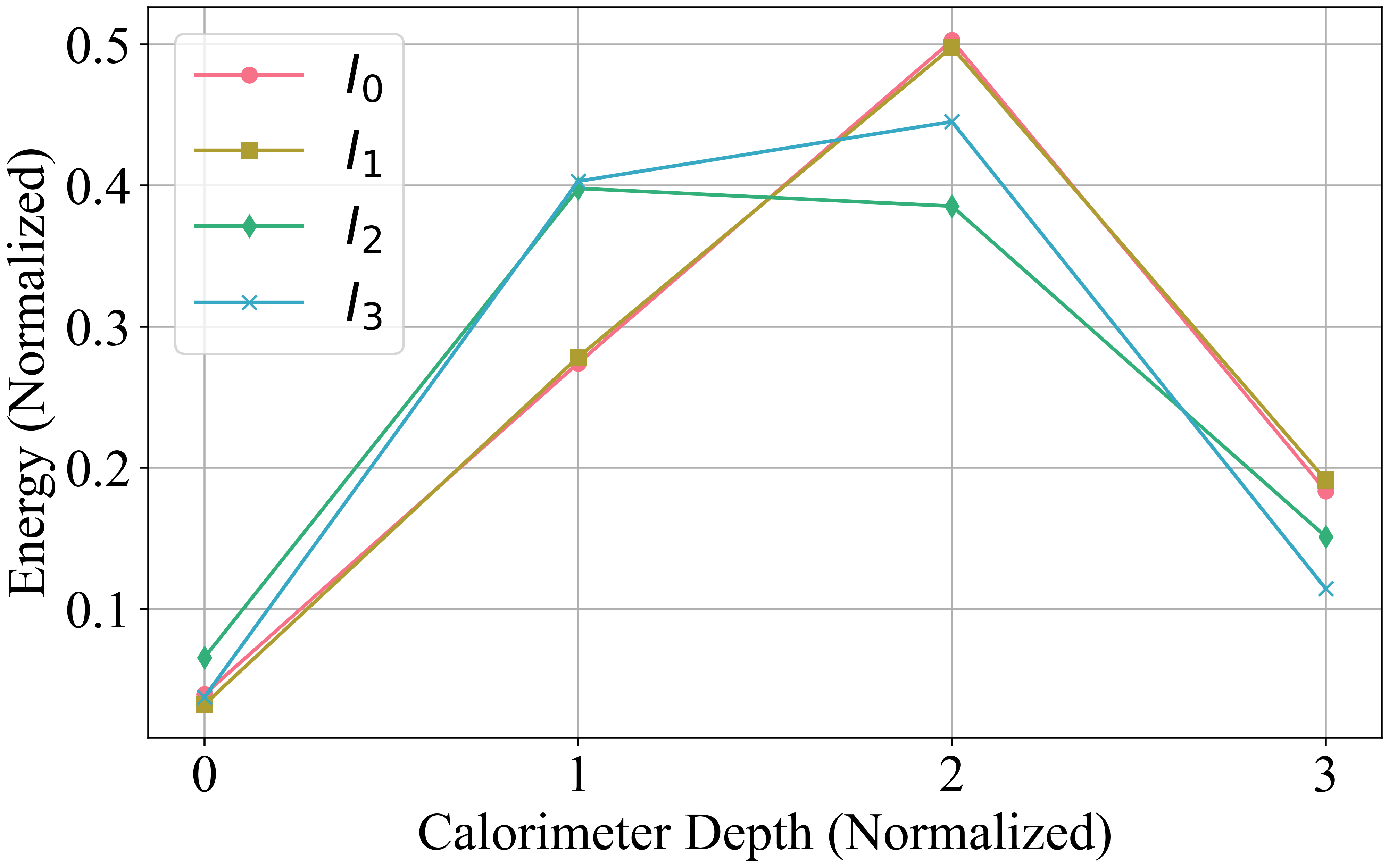}
    \caption{}
    \label{fig:images_perth}
\end{subfigure}
\begin{subfigure}{0.3\textwidth}
    \centering
    \includegraphics[width = \textwidth]{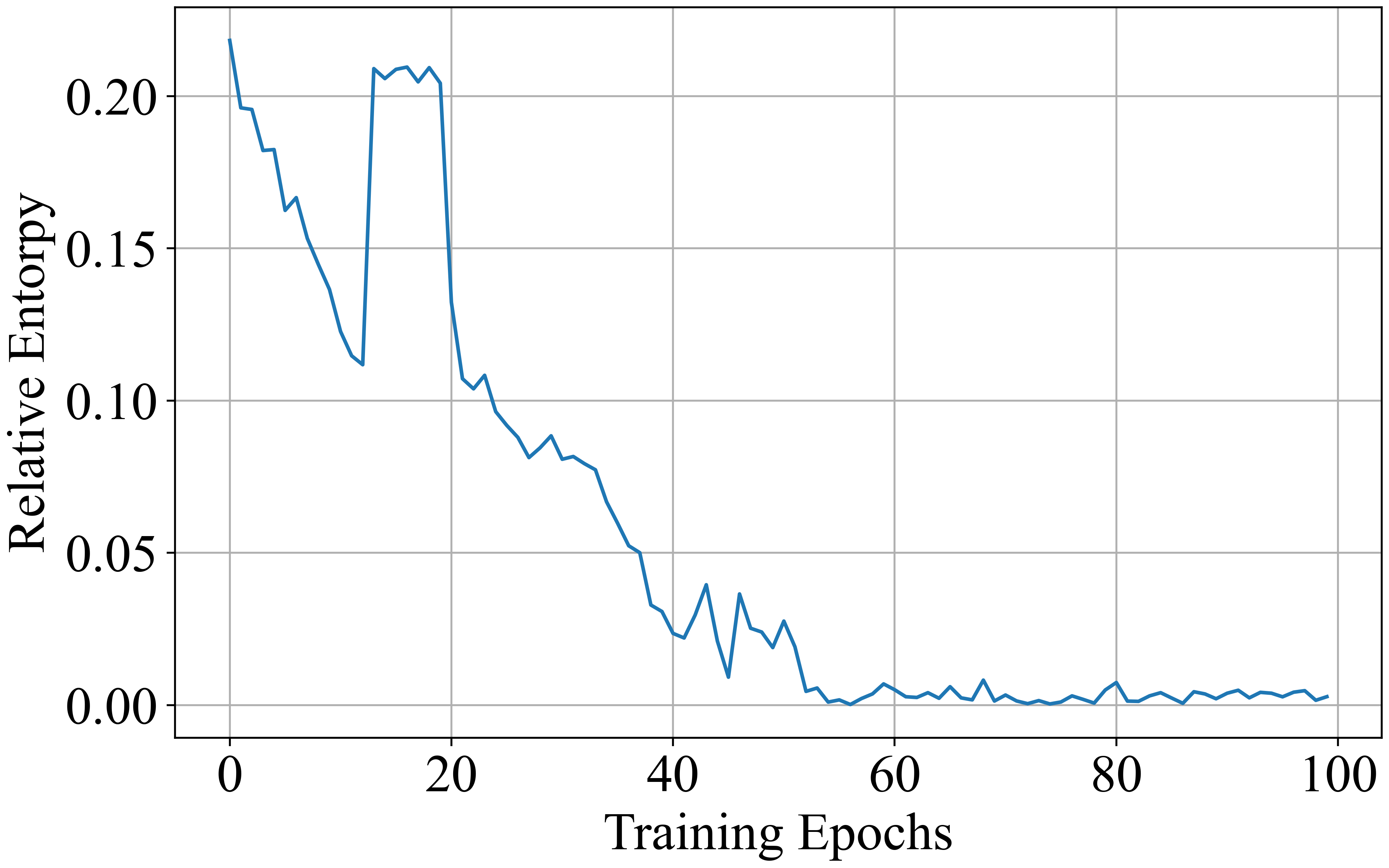}
    \caption{}
    \label{fig:rel_entr_perth}
\end{subfigure}
\caption{The results of dual-PQC GAN training on \textit{ibm\_lagos} (a,b,c) and \textit{ibm\_perth} (d,e,f). The convergence in mean image (a,d) and the decrease in $D_{KL}$ show that the training is successful on average. However, the mode collapse failure is observed for the individual images with an overlap between $\mathcal{I}_2$, $\mathcal{I}_3$ on Figure (b) and $\mathcal{I}_0$, $\mathcal{I}_1$ on Figure (e). }
\label{fig:real_test}
\end{figure}

We conclude the study with the training on real quantum hardware. 
Figure \ref{fig:real_test} displays the result of dual-PQC GAN training on two real IBMQ machines, \textit{ibmq\_lagos} and \textit{ibm\_perth}. The mean images and the progress in relative entropy prove that the dual-PQC model can be successfully trained  \textit{on average} on both devices with final $D_{KL}$ of $7.64\times 10^{-3}$ and $2.74\times 10^{-3}$, respectively. Unfortunately, despite low $D_{KL, ind}$ of $1.29\times 10^{-2}$ and $3.56\times 10^{-3}$, respectively, the current training with real quantum hardware falls into a mode collapse, one of the most common failures in classical GAN~\cite{mode_collapse}, where the model only reproduces a low variety of samples as shown on Figure \ref{fig:images_lagos} and \ref{fig:images_perth}.  

%% file: 5Conclusion.tex
In this paper, we studied the impact of quantum noise on dual-PQC GAN training.  The results showed that with an appropriate choice of hyperparameters, the dual-PQC GAN converges under real working conditions -- specifically in the presence of two-qubit gate noise -- producing images close to the real ones. Moreover, the convergence in mean image is observed on the real quantum hardware training, yet the mode collapse failure emerges as an issue to solve. 

Future research will target methods to avoid the presence of mode collapse, for example by adding an additional term to the loss function. Furthermore, an increase in the size of training set in order to reproduce higher number of images with more pixels will be investigated.